\documentclass[aps,twocolumn,preprintnumbers,superscriptaddress,floatfix,prl,10pt]{revtex4-1}
\usepackage{bbm}
\usepackage{bm}
\usepackage{amsmath}
\usepackage{amssymb}
\usepackage{empheq}
\usepackage{graphicx}
\usepackage{mathrsfs}
\usepackage{amsfonts}
\usepackage{amsthm}
\usepackage{color}
\usepackage{multirow}
\usepackage{bigints}
\usepackage{txfonts}
\usepackage{float}
\usepackage{hyperref}
\hypersetup{
     unicode=false,              
     pdftoolbar=true,           
     pdfmenubar=true,         
     pdffitwindow=false,      
     pdfstartview={FitH},    
     pdftitle={My title},       
     pdfauthor={Author},     
     pdfsubject={Subject},   
     pdfcreator={Creator},   
     pdfproducer={Producer}, 
     pdfkeywords={keyword1} {key2} {key3}, 
     pdfnewwindow=true, 
     colorlinks=false,       
     linkcolor=red,           
     citecolor=green,        
     filecolor=magenta,    
     urlcolor=cyan           
}

\makeatletter \tolerance = 10000 \tolerance = 10000

\makeatother
\begin{document}

\title{Berezinskii-Kosterlitz-Thouless localization-localization transitions in disordered two-dimensional quantized quadrupole insulators}

\author{C. Wang}
\email[Corresponding author: ]{physcwang@tju.edu.cn}
\affiliation{Center for Joint Quantum Studies and Department of Physics, School of Science, Tianjin University, Tianjin 300350, China}
\author{Wenxue He}
\affiliation{Center for Joint Quantum Studies and Department of Physics, School of Science, Tianjin University, Tianjin 300350, China}
\affiliation{Tianjin Key Laboratory of Low Dimensional Materials Physics and Preparing Technology, School of Science, Tianjin University, Tianjin 300072}
\author{Hechen Ren}
\affiliation{Center for Joint Quantum Studies and Department of Physics, School of Science, Tianjin University, Tianjin 300350, China}
\affiliation{Tianjin Key Laboratory of Low Dimensional Materials Physics and Preparing Technology, School of Science, Tianjin University, Tianjin 300072}
\affiliation{Joint School of National University of Singapore and Tianjin University, International Campus of Tianjin University, Binhai New City, Fuzhou 350207, China}
\author{X. R. Wang}
\affiliation{Physics Department, The Hong Kong University of Science 
and Technology (HKUST), Clear Water Bay, Kowloon, Hong Kong}
\affiliation{HKUST Shenzhen Research Institute, Shenzhen 518057, China}

\date{\today}

\begin{abstract}
Anderson localization transitions are usually referred to as quantum phase transitions from delocalized states to localized states in disordered systems. Here we report an unconventional ``Anderson localization transition'' in two-dimensional quantized quadrupole insulators. Such transitions are from symmetry-protected topological corner states to disorder-induced normal Anderson localized states that can be localized in the bulk, as well as at corners and edges. We show that these localization-localization transitions (transitions between two different localized states) can happen in both Hermitian and non-Hermitian quantized quadrupole insulators and investigate their criticality by finite-size scaling analysis of the corner density. The scaling analysis suggests that the correlation length of the phase transition, on the Anderson insulator side and near critical disorder $W_c$, diverges as $\xi(W)\propto \exp[\alpha/\sqrt{|W-W_c|}]$, a typical feature of Berezinskii-Kosterlitz-Thouless transitions. A map from the quantized quadrupole model to the quantum two-dimensional $XY$ model motivates why the localization-localization transitions are Berezinskii-Kosterlitz-Thouless type.
\end{abstract}

\maketitle

\emph{Introduction.$-$}Disorder-induced quantum phase transitions, known as Anderson localization transitions (ALTs)~\cite{pwanderson_pr_1958,bkramer_rpp_1993,bhuckestein_rmp_1995,fevers_rmp_2008}, are a fundamental topoic in wave physics since disorders are ubiquitous in nature and profoundly affect the properties of states, as demonstrated in quantum Hall systems~\cite{bhuckestein_rmp_1995}, topological Anderson insulators~\cite{jli_prl_2009,cwgroth_prl_2009,ysu_prb_2016}, and non-Hermitian systems~\cite{nhatano_prl_1996,yhuang_prb_2020,xluo_prl_2021,cwang_prb_2023}. The scope of ALTs is very broad, including metal-insulator transitions and quantum-Hall-type transitions that occur in topologically non-trivial systems~\cite{fevers_rmp_2008}. They can be either second-order phase transitions between localized and extended states ~\cite{snevangelon_prl_1995,yasada_prl_2002,gorso_prl_2017,cwang_prb_2017} or Berezinskii-Kosterlitz-Thouless (BKT) transitions between localized and critical states with fractal structures~\cite{xiexc_prl_1998,yyzhang_prl_2009,cwang_prl_2015,czchen_prl_2019,wchen_prb_2019}. ALTs are generally considered as  phase transitions from delocalized (extended or critical) states to localized states in disordered systems.
\par

Here we report an unusual disorder-induced phase transition from topologically nontrivial localized states to trivial localized states in two-dimensional quantized quadrupole insulators (2D QQIs)~\cite{jlangbehn_prl_2017,wabenalcazar_science_2017,wabenalcazar_prb_2017,fschindler_sciadv_2018,mezawa_prl_2018}. QQIs are two-dimensional second-order topological insulators with in-gap zero-dimensional corner states characterized by a non-zero quadrupole moment. The second-order topological insulators survive at weak disorders, as genuine topological phases~\cite{cali_prl_2020,hliu_prb_2021,cwang_prresearch_2020,cwang_prb_2021}, in contrast to other localized states in deterministic fractals due to constructively interference \cite{xrw}. Although Anderson localizations are surely dominated at strong disorders, how QQIs evolve into the Anderson insulators (AIs) remains unresolved. Corner densities of wave functions show directly QQIs transition to AIs as disorders increase, different from the delocalization-localization transitions reported in a second-order topological Anderson insulator~\cite{cali_prl_2020}. The correlation lengths diverge exponentially at the AI side near the critical disorder, a feature reminiscent of BKT transitions~\cite{vlberezinskii_spj_1971,jmkosterlitz_jpc_1973}. 
\par

The topological corner-dwelling states of QQIs have been observed in various materials, including sonic~\cite{zzhang_prl_2019,hxue_nc_2020}, photonic~\cite{ktakata_prl_2018,byxie_prl_2019}, cold atomic~\cite{lli_prl_2020}, and magnetic systems~\cite{zxli_pr_2021}, whose Hamiltonians are non-Hermitian in principle. Therefore, we consider a non-Hermitian Hamiltonian whose Hermitian part can be a QQI. Besides, in the presence of non-Hermicity, such models can support a transition from normal insulators in the Hermitian limit to non-Hermitian QQIs. Driven by disorders, both Hermitian and non-Hermitian QQIs undergo localization-localization transitions with the same power-law divergences of correlation lengths, exhibiting the BKT criticality. The BKT criticality can be understood by an equivalent map from the QQIs to the quantum $XY$ model in 2D.
\par

\begin{figure}[htbp]
\centering
\includegraphics[width=0.48\textwidth]{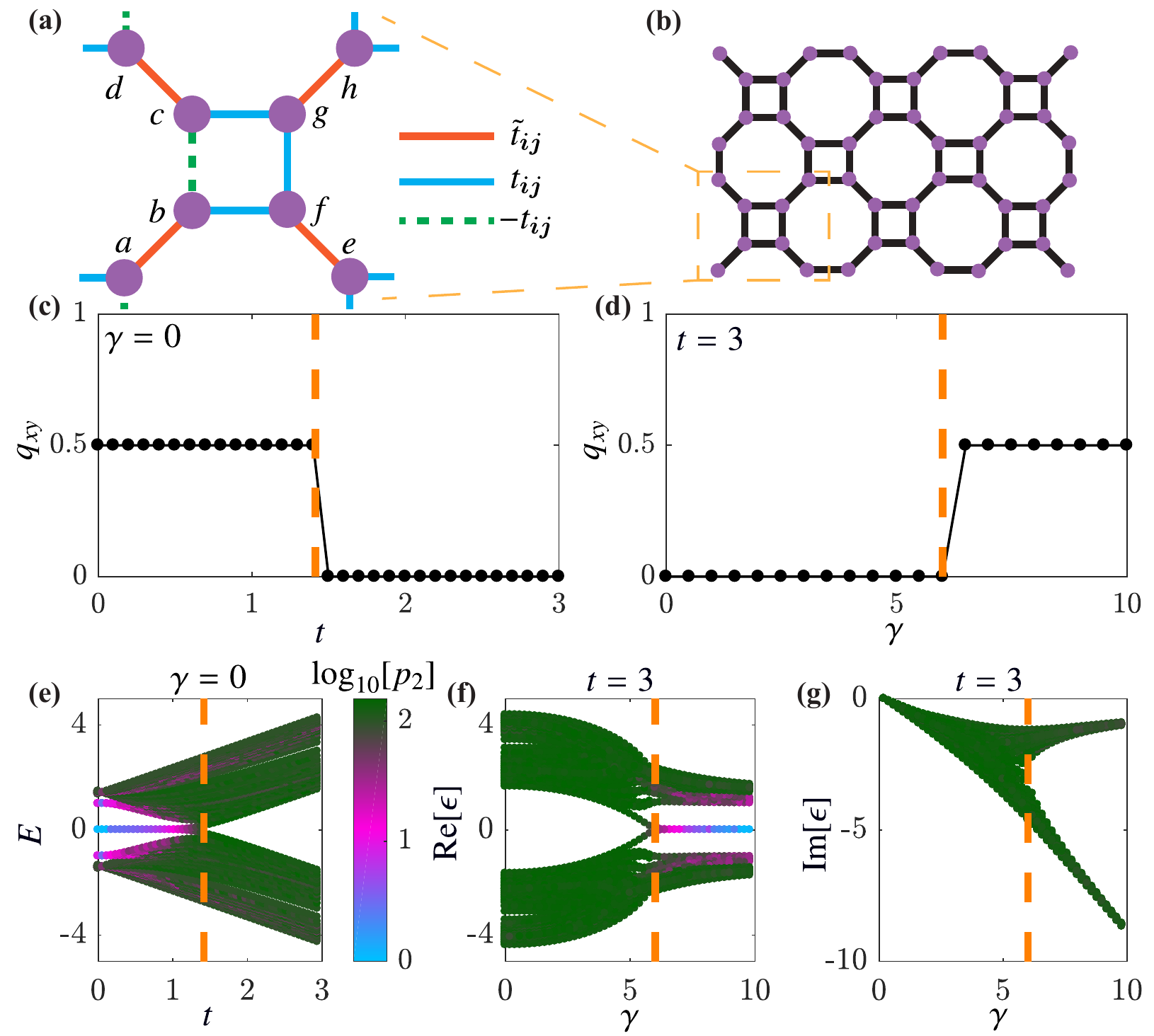}
\caption{(a) Unit cell of the square-octagon lattice. There are three different types of hopping $\tilde{t}_{\bm{ij}}$ (orange), $t_{\bm{ij}}$ (blue), $-t_{\bm{ij}}$ (green), whose amplitudes are random. (b) A rectangle-shape lattice of $L_x=3,L_y=2$ with lattice constant $a=1$. (c) $q_{xy}$ as a function of $t$ for $\gamma=0$ (Hermitian) and $W=0$. (d) $q_{xy}$ v.s. $\gamma$ for $t=3$ and $W=0$. (e) Energy spectrum of $\gamma=0$ for $L_x=L_y=5$ in the clean limit. (f,g) $\text{Re}[\epsilon]$ and $\text{Im}[\epsilon]$ v.s. $\gamma$ for $t=3$ and $W=0$. Colors in (e,f,g) map the common logarithm of participation numbers $\log_{10}[p_2(0)]$ interpreted later. Orange dash lines in (c,d,e,f,g) separate QQIs and normal insulators.}
\label{fig1}
\end{figure}

\emph{Models.$-$}We consider a tight-binding model on a  $L_x\times L_y$ square-octagon lattice subject to a non-Hermitian potential as shown in Figs.~\ref{fig1}(a,b). The Hamiltonian reads
\begin{equation}
\begin{gathered}
H=\sum_{\bm{i}}c^\dagger_{\bm{i}}\epsilon_{\bm{i}}c_{\bm{i}}+\left( \sum_{\bm{ij}}c^\dagger_{\bm{i}}\mathcal{T}_{\bm{ij}}c_{\bm{j}}+h.c.\right).
\end{gathered}\label{eq_1_1}
\end{equation}
Here, $c^\dagger_{\bm{i}}$ and $c_{\bm{i}}$ represent the creation and annihilation operators on site $\bm{i}$. The nearest-neighbor hoppings $\mathcal{T}_{\bm{ij}}$ are, respectively, $\tilde{t}_{\bm{ij}}, t_{\bm{ij}},-t_{\bm{ij}}$ for different types of hoppings shown in Fig.~\ref{fig1}(a), where $\tilde{t}_{\bm{ij}}$ and $t_{\bm{ij}}$ distribute randomly and uniformly in ranges of $[(-W/2+1)\tilde{t},(W/2+1)\tilde{t}]$ and $[(-W/2+1)t,(W/2+1)t]$. Hence, $W$ measures the degree of randomness. Non-Hermiticity is introduced by the on-site potentials with $\epsilon_{\bm{i}}=i\gamma$ ($\gamma\in\mathbb{Re}$) for b, c, g, and f sub-lattices and $\epsilon_{\bm{i}}=0$ otherwise. Below, we set $\tilde{t}=1$ as the energy unit.
\par

\emph{Topological invariance for clean QQIs.$-$}In the clean limit, the Bloch Hamiltonian of model~\eqref{eq_1_1} reads
\begin{equation}
\begin{gathered}
h(\bm{k})=\tilde{t}\tau_1 s_0\sigma_0+(i\gamma/2)\tau_0s_0\sigma_0-(i\gamma/2)\tau_3 s_3 \sigma_0 \\
-\left[ g_+(k_2)\tau_1 s_1-g_-(k_2)\tau_2 s_2 +h(k_2)\left(\tau_2 s_1 +\tau_1 s_2\right)\right]\sigma_3 \\
+\left[ g_+(k_1)\tau_0s_0+g_-(k_1)\tau_3s_3 \right]\sigma_1+h(k_1)\left( \tau_0 s_0+\tau_3 s_3 \right)\sigma_2.
\end{gathered}\label{eq_1_2}
\end{equation}
$\{\tau_{i=0,1,2,3}\},\{ s_{i=0,1,2,3}\},\{\sigma_{i=0,1,2,3}\}$ are unit and Pauli matrices on the spaces $\{ \Psi^\dagger_{a,\bm{k}}, \Psi^\dagger_{b,\bm{k}},\Psi^\dagger_{c,\bm{k}},\Psi^\dagger_{d,\bm{k}},\Psi^\dagger_{e,\bm{k}},\Psi^\dagger_{f,\bm{k}},\Psi^\dagger_{g,\bm{k}},\Psi^\dagger_{h,\bm{k}}\}$. $g_{\pm}(k)=(t/2)(\cos[k]\pm 1),h(k)=(t/2)\sin [k]$. Under certain conditions, $h(\bm{k})$ is a QQI for both Hermitian ($\gamma=0$) and non-Hermitian ($\gamma\neq 0$) cases, where topologically-protected states appear at the corners of a finite-size lattice like Fig.~\ref{fig1}(b). 
\par

Such corner states are featured by a bulk topological quadrupole moment $q_{xy}$, which can be calculated by the nested Wilson loops~\cite{wabenalcazar_prb_2017}. A Wilson loop operator in the $x$ ($y$) direction is defined as $\hat{\mathcal{W}}_{x(y)}$. In the Hermitian limit, the matrix elements of the Wilson-loop operator in the $y$ direction are given by a path-integral over the first Brillouin zone, denoted by $\overline{\exp}$,
\begin{equation}
\begin{gathered}
\left[ \hat{\mathcal{W}}_{y}(\mathbb{C}_i)\right]_{mn}=\overline{\exp}\left[ i\oint_{\mathbb{C}_i}A^{y}_{mn}(k_1,k_2)dk_2 \right]
\end{gathered}\label{eq_1_3}
\end{equation}
with $\mathbb{C}_i$ being a closed loop of $k_1=q_i\Delta k_1$ and $k_2=0\to \Delta k_2 \to 2\Delta k_2\cdots\to (L_y-1)\Delta k_2\to 0$ and $\Delta k_1=2\pi/L_x, \Delta k_2=2\pi/L_y$. $A^{y}_{mn}(k_1,k_2)=i\langle u_m(\bm{k})|\partial_{k_2}|u_n(\bm{k})\rangle$ is the Berry connection, and $|u_n(\bm{k})\rangle$ is the occupied Bloch function~\cite{gauge,tfukui_jpsj_2005}. By diagonalizing the Wilson-loop operator under periodic boundary conditions on both directions, one can define the Wannier Hamiltonian $\hat{\mathcal{H}}_{y}$ as $\hat{\mathcal{W}}_{y}=e^{ i\hat{\mathcal{H}}_{y}}$. We diagonalize the Wilson-loop operator $\hat{\mathcal{W}}_{y}|v^q_y(k_1)\rangle=e^{i2\pi v^q_y(k_1)}|v^q_y(k_1)\rangle$ and calculate the Wannier bands as $|w^q_y(\bm{k})\rangle=\sum^{N_{\text{occ}}}_{n=1}\left[ v^q_y(k_1) \right]^n |u_n(\bm{k})\rangle$ with $\left[ v^q_y(k_1) \right]^n$ being the $n$th components of the $q$th eigenket $|v^q_y(k_1)\rangle$. Here, $N_{\text{occ}}$ is referred to as the number of occupied bands. The associated polarization in the subspace of Wannier bands is given by $p^q_x=-1/(2\pi)\int_{\text{BZ}}\tilde{A}^q_{x}(\bm{k})d^2k$, where $\tilde{A}^q_{x}(\bm{k})=i\langle w^q_y(\bm{k})|\partial_{k_1}|w^q_y(\bm{k})\rangle$ is  the Berry connection defined in the Wannier bands. Likewise, one can define the polarization in the $y$ direction by changing $x$ to $y$ ($k_1$ to $k_2$).
\par

A non-zero polarization in the $x(y)$ direction implies that the Wannier Hamiltonian $\hat{\mathcal{H}}_{x(y)}$ is a first-order topological insulator if the system is cut parallel to the $x(y)$ direction. Naturally, the topological quadrupole moment is defined as $q_{xy}=2p_xp_y$. A non-zero $q_{xy}$ guarantees a second-order topological insulator with corner states, if open boundary conditions are applied in both directions. Here, $p_{x(y)}=\sum_q p^q_{x(y)}$. Figure~\ref{fig1}(c) shows the numerically obtained $q_{xy}$ as a function of $t$. It is seen that $q_{xy}=0.5$ and 0 for $t<\sqrt{2}$ and $t>\sqrt{2}$. Hence, a topological phase transition from QQIs to normal insulators happens at $t=\sqrt{2}$, where the bulk-gap closes. This is further supported by the energy spectrum plot of model~\eqref{eq_1_1} on a square sample of $L_x=L_y=5$, see Fig.~\ref{fig1}(e). 
\par

Remarkably, non-Hermitian potentials can drive a quantum phase transition from normal insulators to QQIs, even when the system is topologically-trivial in the Hermitian limit~\cite{ytian_prb_2023}. We generalize the Wilson-loop operators to non-Hermitian systems by rendering the Berry connection as $A^{x(y)}_{mn}=i\langle u^{\text{L}}_n(\bm{k})|\partial_{k_1(k_2)}|u^{\text{R}}_m(\bm{k})\rangle$ with $|u^{\text{R}}_m(\bm{k})\rangle$ and $|u^{\text{L}}_m(\bm{k})\rangle$ being the right and left Bloch kets, respectively, i.e., $h(\bm{k})|u^{\text{R}}_m(\bm{k})\rangle=\epsilon_m(\bm{k})|u^{\text{R}}_m(\bm{k})\rangle$, $h^\dagger(\bm{k})|u^{\text{L}}_m(\bm{k})\rangle=\epsilon^\ast_m(\bm{k})|u^{\text{L}}_m(\bm{k})\rangle$, and $\langle u^{\text{L}}_m(\bm{k})|u^{\text{R}}_n(\bm{k})\rangle=\delta_{mn}$~\cite{xwluo_prl_2019}. The numerically calculated $q_{xy}$ as a function of $\gamma$ for $t=3$ is plotted in Fig.~\ref{fig1}(d). Similar to the Hermitian cases,  QQIs appear for $\gamma>6$, which is also consistent with calculations of the complex energy spectrum of $H$, see Figs.~\ref{fig1}(f,g). 
\par

\emph{Hermitian systems.$-$}Having illustrated the existence of QQIs in the clean limit, we now investigate the disorder-induced quantum phase transitions. Consider Hermitian QQIs first. Model~\eqref{eq_1_1} preserves both time-reversal symmetry (TRS) and particle-hole symmetry (PHS) with symmetry operators $U_{\mathcal{T}}=\tau_0 s_0\sigma_0\otimes I$ and $U_{\mathcal{P}}=\tau_3s_0\sigma_3\otimes I$, respectively. (Here, $I$ is the unit matrix acting on the coordinate space.) Therefore, model~\eqref{eq_1_1} of $\gamma=0$ belongs to class BDI according to the Altland-Zirnbauer classification~\cite{aaltland_prb_1997}.  
\par

\begin{figure}[htbp]
\centering
\includegraphics[width=0.48\textwidth]{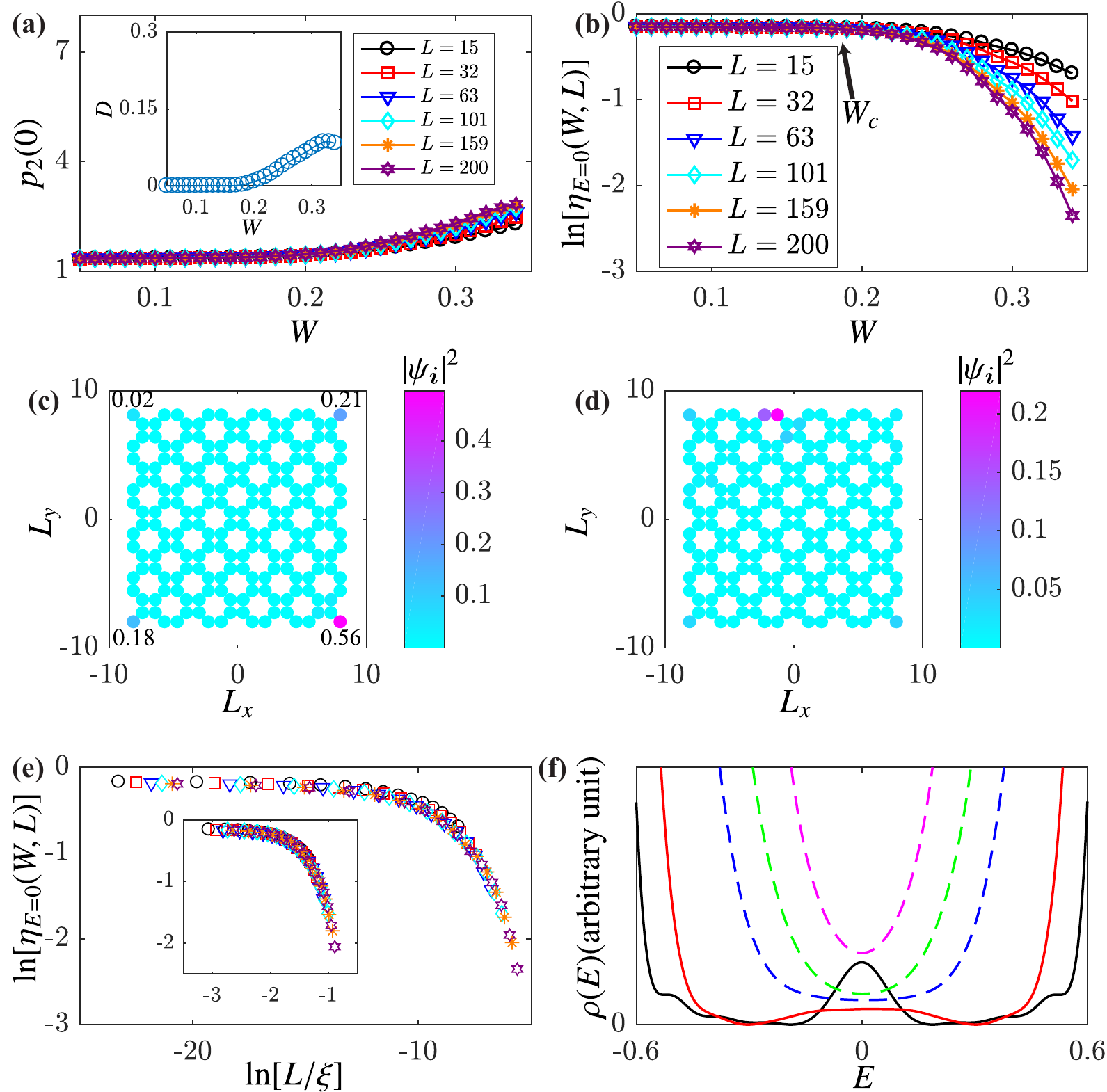}
\caption{(a) $p_2(0)$ as a function of $W$ for the zero-energy state. Inset: Wave-function dimension $D$ as a function of $W$. (b) $\ln[\eta_{E=0}(W,L)]$ v.s. $W$. Arrow locates $W_c$ obtained by the finite-size scaling analyses. (c) Spatial distribution $|\psi_{\bm{i}}|^2$ for $W=0.05$ in a typical realization. $|\psi_{\bm{i}}|^2$ for the four corners are marked. (d) Same as (c) but for $W=0.35$. (e) Scaling function $\ln[\eta_{E=0}(W,L)]=f(\ln [L/\xi])$ with $\xi\propto \exp[\alpha/\sqrt{|W-W_c|}]$. Inset: Same as (e) but $\xi\propto|W-W_c|^{-\nu}$. (f) Average DOS $\rho(E)$ (arbitrary unit) for $W=0.02,0.1,0.2,0.25,0.3$. Those above (below) $W_c$ are dashed (solid) lines. }
\label{fig2}
\end{figure}

For $t<\sqrt{2}$ where $q_{xy}=0.5$, we first compute the  participation number $p_2(E)=\langle \sum_{\bm{i}}|\psi_{\bm{i}}(E)^4|\rangle^{-1}$ for a state of energy $E$ and amplitude $\psi_{\bm{i}}(E)$, which measures how many sites the state occupies~\cite{numerical,kwant,scipy}. It scales with the system size $L_x=L_y=L$ as $p_2\propto L^D$ with $D=d$ (embedded spatial dimensionality), $D<d$, and $D=0$ for extended, critical, and localized states, respectively~\cite{xrwang_pra_1989}. 
\par

Figure~\ref{fig2}(a) exemplifies $p_2(0)$ as a function of $W$ for $t=0.5$ and $L$ ranging from 15 to 200. It is seen that $p_2(0)<3$ until $L=200$. The dimension $D(W)$, obtained by a log-log plot of $L$ and $p_2(0)$, is almost 0 ($D<0.1$, see inset of Fig.~\ref{fig2}(a)). These features suggest that zero-energy states are localized~\cite{p2_fractal}. This is the main difference between our model and that of Ref.~\cite{cali_prl_2020}, where delocalized states appear near topological phase transition points, even though both belong to class BDI. Indeed, our numerical calculations are consistent with the one-parameter scaling theory of localization which predicts extended states are prohibited in the presence of TRS and spin-rotational symmetry ($U^2_{\mathcal{T}}=I$)~\cite{palee_rmp_1985}. We further substantiate that all states of our model are localized by analysing the dimensionless conductance obtained using the Landauer formalism~\cite{amackinnon_zpb_1985} (see Supplementary~\cite{supp}). 
\par

As $p_2(0)$ is not a good scaling variable to identify the quantum phase transitions, we use the corner density $\eta_{E=0}(W,L)$ to distinguish corner states from conventional localized states. $\eta_{E=0}(W,L)=\sum_{\bm{i}\in \text{corners}}|\psi_{\bm{i}}(0)|^2$, where the summation is over the four corner sites of Fig.~\ref{fig1}(b) for given disorder $W$ and size $L$. Since $\eta_{E=0}(W,L)$ describes the wave-function distribution of the zero-energy states on corners, it should approach a finite non-zero constant for corner states and is zero for topologically trivial (both localized and extended) states. These features can be used to see whether a state of $E$ is a corner state or not~\cite{cwang_prresearch_2020,cwang_prb_2021}.
\par

We numerically calculate the ensemble-average corner density $\eta_{E=0}(W,L)$ for various $W$ and $L$ ranging from 15 to 200, as shown in Fig.~\ref{fig2}(b). Clearly, there exists a critical disorder $W_c$ separating two different regimes. (i) For $W<W_c$,  $\eta_{E=0}(W,L)\simeq 0.86$, and $d\eta_{E=0}(W,L)/dL=0$, indicating the zero-energy states are corner states. (ii) For $W>W_c$, $\eta_{E=0}(W,L)$ decreases with $L$. This feature, together with the $p_2(0)$ results in Fig.~\ref{fig2}(a), suggests that the zero-energy states are conventional localized states (at corners, edges, or bulk). The wave-function distributions for two typical disorders of $W=0.05$ and $W=0.25$ are plotted in Figs.~\ref{fig2}(c,d), which are consistent of our analysis above.  
\par

We have seen a transition from a QQI to an AI at $W_c$ in Fig.~\ref{fig2}(b) and expect $\eta_{E=0}(W,L)$ on the AI's side satisfying the scaling hypothesis~\cite{cwang_prl_2015}
\begin{equation}
\begin{gathered}
\eta_{E=0}(W,L)=f(L/\xi)+CL^{-y}
\end{gathered}\label{eq_1_6}
\end{equation}
where $\xi(W)$ is the correlation length, $f(x)$ is the universal scaling function, $C$ is a constant, and $y>0$ is the exponent for the irrelevant scaling variable. At $W_c$, the correlation length diverges.
\par

To substantiate the single-parameter scaling hypothesis, we show in Fig.~\ref{fig2}(e) and its inset that all curves of $\ln[ \eta_{E=0}(W,L)]$ near $W_c$ collapse into a single smooth curve $f(\ln [L/\xi])$ if proper $\xi$ are chosen and the effect of irrelevant scaling variables is removed~\cite{wchen_prb_2019}. If $\xi$ is assumed to follow a power law, $\xi\propto|W-W_c|^{-\nu}$, critical exponent $\nu=5.9\pm 0.6$, which is higher than any known critical exponent in 2D systems, is obtained. However, if the localization-localization transition from QQIs to AIs is assumed to be BKT-type and $\xi$ exponentially diverge as $\xi\propto\exp[\alpha/\sqrt{|W-W_c|}]$, a good scaling function with $\alpha=4.5\pm 0.5$ and $W_c=0.17\pm 0.01$ and an acceptable goodness-of-fit $Q=0.3$ are obtained. 
\par

The origin of the quantum phase transition in Fig.~\ref{fig2}(b) is revealed by the average density of states (DOS) plot, defined as $\rho(E)=\langle\sum_q \delta(E-E_q) \rangle/L^2$, as shown in Fig.~\ref{fig2}(f). $\rho(E)$ is obtained from the kernel polynomial method~\cite{aweise_rmp_2006}. The DOS is symmetric about $E=0$ due to PHS. For $W<W_c$, we see a gap of bulk states near $E=0$, which becomes smaller with the increase of $W$ and disappear when $W>W_c$. The in-gap peak that stands for corner states also disappear when $W>W_c$. Hence, such topologically quantum phase transition is a gap-closing transition, similar to those in three-dimensional second-order topological insulators~\cite{cwang_prresearch_2020} and Weyl semimetals~\cite{sliu_prl_2016}.
\par 

\emph{Non-Hermitian systems.$-$}BKT transitions also appear in non-Hermitian QQIs. For $\gamma=10$ and $t=3$, model~\eqref{eq_1_1} is a non-Hermitian QQI with corner states at $\text{Re}[\epsilon]=0$, see Fig.~\ref{fig1}(f). Now, we generalize the participation numbers and the corner density by replacing $\psi_{\bm{i}}(E=0)$ to the right eigenfunction $\psi^{\text{R}}_{\bm{i}}(\text{Re}[\epsilon]=0)$. 
\par

\begin{figure}[htbp]
\centering
\includegraphics[width=0.48\textwidth]{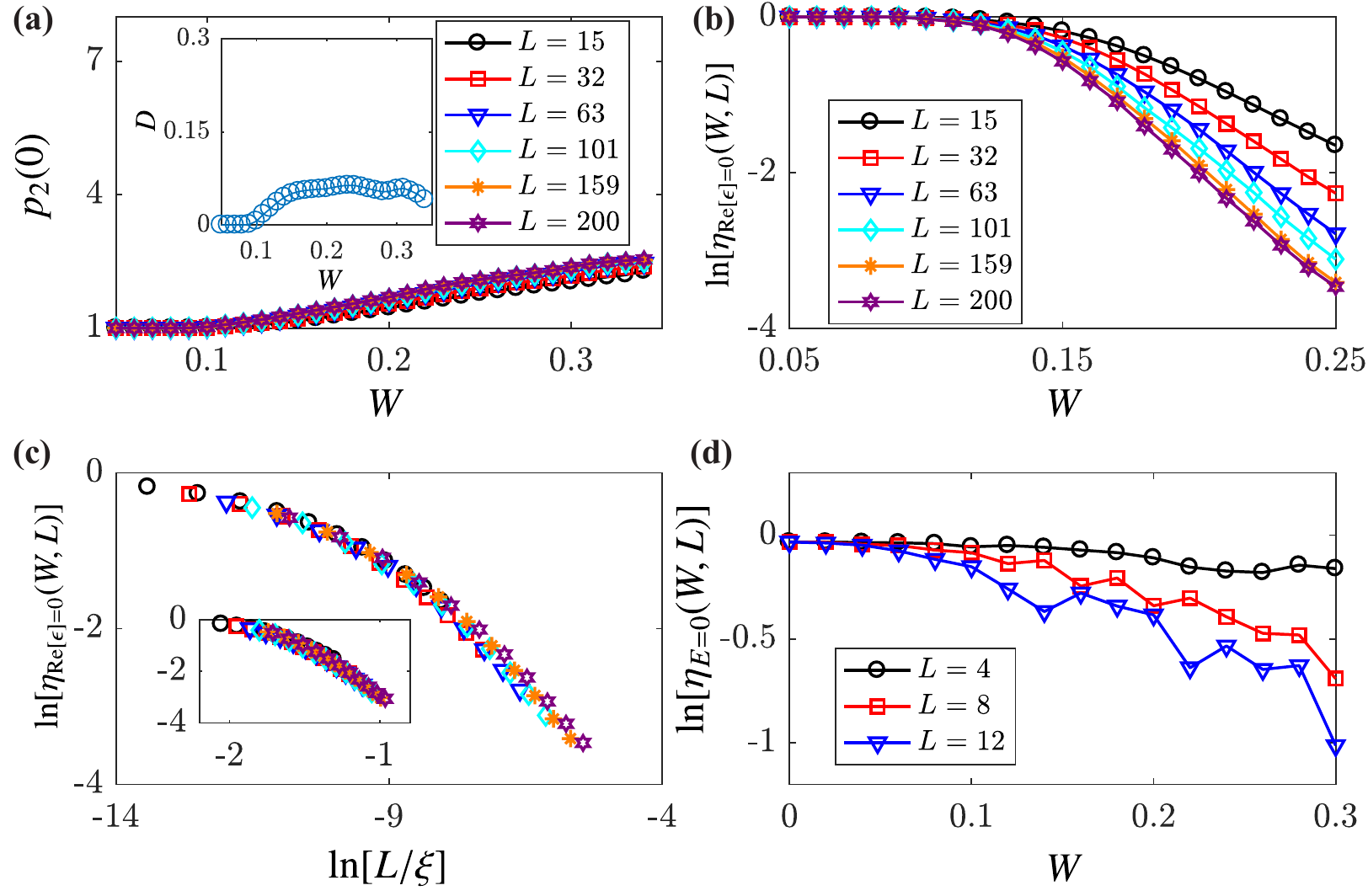}
\caption{(a) $p_2(0)$ as a function of $W$ for the $\text{Re}[\epsilon]=0$ state. Inset: $D(W)$ for $\text{Re}[\epsilon]=0$. (b) $\ln[\eta_{\epsilon}]$ v.s. $W$ for the corner states and different $L$. (c) Scaling function $\ln[\eta_{\epsilon=0}(W,L)]=f(\ln [L/\xi])$ with $\xi\propto\exp [\alpha/\sqrt{|W-W_c|}]$. Inset: Same as (c) but for $\xi\propto |W-W_c|^{-\nu}$. (d) $\eta_{E=0}(W,L)$ data from circuit simulation for $L=4,8,12$. The system is a QQI when $W=0$.}
\label{fig3}
\end{figure}

Figure~\ref{fig3}(a) and its inset display $p_2(0)$ and $D$ as a function of $W$ for $\gamma=10$ and $t=3$, from which one can see all states are also localized. Model~\eqref{eq_1_1} with $\gamma\neq 0$ preserves TRS$^\dagger$, PHS$^\dagger$, and chiral symmetry with symmetry operators $U_{\mathcal{P}+}=\tau_0 s_0\sigma_0\otimes I$, $U_{\mathcal{T}-}=\tau_0 s_0\sigma_3\otimes I$, and $U_{\mathcal{C}}=U_{\mathcal{P}+}U_{\mathcal{T}-}$ and belongs to class BDI$^\dagger$ of the non-Hermitian Altland-Zirnbauer classification~\cite{kkawabata_prx_2019}. A theory by Luo et al. predicts an equivalence of criticality between class BDI$^\dagger$ and class AI in Hermitian systems~\cite{xluo_prresearch_2022}, where all states are localized in 2D. Our calculations are consistent with it.
\par

Although there is no delocalization-localization transition, a localization-localization transition from QQIs to AIs can be seen from the plot of $\eta_{\text{Re}[\epsilon]=0}$, where $d\eta_{\text{Re}[\epsilon]=0}/dL=0$ and $<0$ for $W\leq W_c=0.08\pm 0.01$ and $W>W_c$, respectively. One can choose a proper correlation length $\xi(W)$ such that $\eta_{\text{Re}[\epsilon]=0}$ on the AI side merge into a single scaling function $f(x)$. The correlation length $\xi$ diverges either in a power law $\xi\propto |W-W_c|^{-\nu}$ with $\nu=7.61\pm 0.02$ or an exponential decay $\xi\propto\exp[\alpha/\sqrt{|W-W_c|}]$ with $\alpha=4.56\pm 0.04$, see Figs.~\ref{fig3}(c,d). Since $\nu$ is very large (generally speaking, $\nu<3$ in 2D non-Hermitian systems, see Ref.~\cite{xluo_prresearch_2022}), we thus argue that the transition in Fig.~\ref{fig3}(b) is BKT-type.
\par

\emph{Experimental relevance.$-$}We have designed an LC circuit which is equivalent to model~\eqref{eq_1_1} in the Hermitian limit at resonance frequency (see Supplementary~\cite{supp}). The corner density is now measurable: $\eta_{E=0}(W,L)=\sum_{\bm{i}\in\text{corners}}|Z_{\bm{i}}|/Z$ with $Z_{\bm{i}}$ and $Z$ being the ground impedance of the four corner sites and the total impedance, respectively. We use LTspice~\cite{ltspice}, a well-established electric circuit simulator, to calculate $\ln[\eta_{E=0}(W,L)]$ as a function of $W$ for $L=4,8,12$, see Fig.~\ref{fig3}(d), which exhibits similar behaviors to Fig.~\ref{fig2}(b) and serves as strong evidence of the localization-localization transitions.
\par

\emph{Mapping QQIs to quantum $XY$ model.$-$}We now try to understand why the localization-localization transitions are BKT-type. When $W<W_c$, the zero-energy states are highly localized at four corners. The effective Hamiltonian of the four corner states of $E=0$ is $\mathbb{H}=\sum_{\bm{i}}\epsilon'_{\bm{i}}c^\dagger_{\bm{i}}c_{\bm{i}}+\sum_{\langle \bm{ij}\rangle}t' c^\dagger_{\bm{i}}c_{\bm{j}}$, here $c^\dagger_{\bm{i}}$ ($c_{\bm{i}}$) are the creation (annihilation) operator of the corner state at site $\bm{i}$ with energy $\epsilon'_{\bm{i}}$ fluctuating around $0$ due to its environment. $t'$ is the coupling between two corner states which depends on the overlap of their wavefunctions. 
\par

Using 2D Wigner-Jordan transformation~\cite{yrwang_prb_1992}, $c^\dagger_{\bm{i}}=e^{i\theta_{\bm{i}}}S^+_{\bm{i}},c_{\bm{i}}=e^{-i\theta_{\bm{i}}}S^-_{\bm{i}}$ with $S^{\pm}_{\bm{i}}=S^{x}_{\bm{i}}\pm i S^y_{\bm{i}}$, to change the Fermion operators $c^\dagger_{\bm{i}}$, $c_{\bm{i}}$  into spin operators $S_{i}^{x}, S_{i}^{y}, $ and $S_{i}^{x}$, Hamiltonians becomes $\mathbb{H}=\sum_{\bm{i}}\epsilon'_{\bm{i}}(S^z_{\bm{i}}+1/2 )+\left( t'\sum_{\langle\bm{ij}\rangle}S^+_{\bm{i}}e^{i(\theta_{\bm{i}}-\theta_{\bm{j}})}S^-_{\bm{j}}+h.c. \right)$. In the transformation, $\theta_{\bm{i}}=\sum_{\bm{j}\neq\bm{i}}c^\dagger_{\bm{i}}c_{\bm{i}}\text{Im}[\ln (z_{\bm{i}}-z_{\bm{j}})]$. $z_{\bm{i}}=x_{\bm{i}}+i y_{\bm{i}}$ is the complex-coordinate of site $\bm{i}$. Within the mean-field approximation~\cite{xiexc_prl_1998}, $\langle \epsilon'_{\bm{i}}\rangle=0$ and 
\begin{equation}
\begin{gathered}
\mathbb{H}= t'\sum_{\langle\bm{ij}\rangle}S^+_{\bm{i}}S^-_{\bm{j}}+h.c.,
\end{gathered}\label{eq_1_9}
\end{equation}
where a constant phase of $\pi/2$ is gauged away as shown in Ref.~\cite{xiexc_prl_1998}.
\par

Equation~\eqref{eq_1_9} is the 2D quantum $XY$ model that supports a BKT transition characterized by exponentially divergent correlation lengths and in-plane susceptibility~\cite{hqding_prb_1990}. This naturally explains why the reported localization-localization transitions at $E=0$ are BKT-type. We would like to make the following remarks. (i) The mean-field approximation works well near $W_c$, on both QQIs and AIs sides, where the wave-function overlaps of corner states are not small. This can be seen in Fig.~\ref{fig2}(c): $\psi^2_{\bm{i}}(E=0)$ are finite at lower left and upper right corners even it is maximal at the lower right corner. (ii) The above map is based on the wave-function nature of corner states and should apply to non-Hermitian QQIs as well where $\langle\epsilon'_{\bm{i}}\rangle$ are complex. Hence, BKT transitions are also expected in non-Hermitian QQIs, see Fig.~\ref{fig3}.
\par

\emph{Summary.$-$}In summary, we have shown that 2D QQIs, both Hermitian and non-Hermitian, can undergo disorder-driven localization-localization transitions, whose criticality is BKT-type. The BKT-type localization-localization transitions of corner states can also be seen in second-order topological insulators characterized by $Z_3$ Berry phases~\cite{mezawa_prl_2018} (see Supplementary~\cite{supp}). Although our claim of BKT-type localization-localization transitions agrees with the field-theoretical approach on BDI class (same as our Hermitian QQIs)~\cite{ejkonig_prb_2012}, whether the BKT transitions can happen for class BDI$^\dagger$ (non-Hermitian QQIs) remains unclear. Finally, BKT transitions are from ordered topologically localized states like bound vortex-anti-vortex pairs in the 2D XY model to conventionally localized states like unpaired vortices and anti-vortices~\cite{vlberezinskii_spj_1971,jmkosterlitz_jpc_1973}. Intuitively, the localization-localization transitions, from topologically in-gap states bounded to the corners to conventional localized states, should also be BKT-type due to the similarities with the original vortex BKT transitions.
\par

\begin{acknowledgments}

This work is supported by the National Natural Science Foundation of China (Grants No.~11774296, No.~11704061, and No.~11974296) and Hong Kong RGC (Grants No.~16300522 and No.~16302321).

\end{acknowledgments}


\begin{thebibliography}{99}
\bibitem{numerical}
Numerically, we compute the participation numbers and the corner densities by using the KWTANT package~\cite{kwant} and the SciPy library~\cite{scipy} on Python.
\end{thebibliography}
\end{document}